\documentclass[twocolumn,prb,superscriptaddress]{revtex4}
\usepackage{graphicx}
\newcommand{\be}{\begin{equation}}
\newcommand{\ee}{\end{equation}}
\newcommand{\bea}{\begin{eqnarray}}
\newcommand{\eea}{\end{eqnarray}}

\newcommand{\p}{\partial}

\newcommand{\la}{\langle}
\newcommand{\ra}{\rangle}

\newcommand{\lp}{\left(}
\newcommand{\rp}{\right)}

\renewcommand{\phi}{\varphi}
\renewcommand{\epsilon}{\varepsilon}
\renewcommand{\vec}[1]{{\bf #1}}
\renewcommand{\Im}{{\rm Im}\,}
\renewcommand{\Re}{{\rm Re}\,}

\begin{document}

\title{Spatial Ordering of Defects and Conductivity of Functionalized Graphene}
\author{D. A. Abanin}
\affiliation{Princeton Center for Theoretical Science and Department of Physics, Princeton University, Princeton NJ 08544, USA}
\author{L. S. Levitov}
\affiliation{Department of Physics, Massachusetts Institute of Technology, Cambridge MA 02139, USA}
\begin{abstract}
Recently, it was predicted that an RKKY-type interaction between adatoms in graphene can drive an ordering transition to a state with broken sublattice symmetry (arXiv:1004.3678). In this state, due to Bragg scattering of electron waves on the sublattice modulation, a gap opens up at the Dirac point (DP). Here we investigate the effect of ordering on the transport properties, finding that upon transition from disordered state to an ordered state the conductivity is increased (reduced) at energies away (near) DP. This behavior can be understood as a result of coherent Bragg scattering in the ordered state. The conductivity change can serve as a direct signature of adatom ordering.  
\end{abstract}

\maketitle

\section{Introduction}

Electronic properties of graphene can change in dramatic ways in the presence of adatoms or chemical groups, such as H, F or OH, which bind to carbon atoms via $sp^2\to sp^3$ hybridization, effectively turning the functionalized carbon atoms into vacancies. Microscopically, such defects represent resonant scatterers for the conduction electrons, with the resonance energy positioned near the Dirac point (DP) \cite{Wehling08}. Resonant enhancement of scattering at low energies has dramatic effect on electronic properties. In recent experiments on hydrogenated graphene \cite{Elias09,Rotenberg10} it was found that even minute amounts of hydrogen can make the resistance near DP very high, turning the temperature dependence from metal-like to semiconductor-like. This behavior demonstrates that adatoms on graphene offer a new knob to control electronic properties of this material. 

Properties of adatoms on metal surfaces, and on graphene, are strongly affected by the RKKY-type interactions mediated by the conduction electrons\cite{Einstein73,Lau78,Einstein78,Cheianov06}. The RKKY interaction between adatoms in graphene was predicted to be strongly enhanced due to the resonant character of scattering at low energies\cite{Shytov09}. In addition, the sign of interaction was found to be sublattice dependent, being different for the atoms residing on the same or on different sublattices, $A$ or $B$. Such sublattice dependence indicates that the RKKY interaction may drive a transition to an ordered state in which the adatoms preferentially populate the $A$ or $B$ sites \cite{Abanin10,Cheianov10}. Bragg scattering of electron waves on the sublattice modulation results in band gap opening at DP, stabilizing the ordered state by a Peierls-type mechanism. The spectral gap was found to be nearly free of localized states as long as the adatoms reside preferentially on sites of one type, $A$ or $B$, even when their distribution is otherwise random.

Here we shall discuss how such ordering manifests itself in transport properties, focusing on the metallic state realized far from DP. 
In our analysis of resonant scattering on the adatoms we use the T-matrix approach, similar to that used to describe scattering on vacancies\cite{Ostrovsky06}. The system with the defects randomly distributed over sublattices $A$ and $B$  are treated within the self-consistent T-matrix approximation (SCTA), in which we perform a virial expansion in the adatom concentration $n_A$, $n_B$. The contribution to the conductivity which depends on the sublattice occupancy fraction $n_A/n_B$ arises at subleading order in the expansion in powers of $n_A$ and $n_B$, see Eq.(\ref{eq:tildePiRA2}) below. Our virial expansion is valid in the metallic regime, at densities which are detuned far enough from neutrality, such that
\be\label{eq:k_Fl_log}
k_F\ell \gg \ln\frac{W}{|\epsilon|}
\ee
with $k_F$ the Fermi momentum and $\ell$ the mean free path. 

The contribution to conductivity $\Delta\sigma$ which depends on $n_A/n_B$ exhibits different behavior at energies near DP and far from it. Near DP we find that ordering results in a decrease in conductivity, $\Delta\sigma<0$, which is consistent with the picture of an energy gap opening at $\epsilon>0$ near DP upon ordering\cite{Abanin10}. The contribution $\Delta\sigma$ changes sign at an energy $\epsilon\approx
0.5\delta$, where $\delta$ is an energy parameter describing the resonance structure of a T-matrix [see Fig.\ref{fig1} and Eq.(\ref{eq:T})]. 

In the metallic regime realized at high detuning from DP, $|\epsilon|\gg|\delta|$, we find a positive contribution which exhibits a universal behavior
\be\label{eq:delta_sigma}
 \Delta\sigma (\epsilon)=\frac{e^2}{\pi^2} \frac{({n}_A-{n}_B)^2}{({n}_A+{n}_B)^2}
\ln \frac{W}{|\epsilon|}
,
\ee
where $W$ is a bandwidth parameter [see Eq.(\ref{eq:T})]. 
The positive sign of $\Delta\sigma$ and its dependence on $n_A/n_B$ mean that this contribution makes the net conductivity higher at $n_A\ne n_B$ than at $n_A=n_B$.
An increase in conductivity due to ordering, $\Delta\sigma>0$, may seem somewhat surprizing given that the ordered state is characterized by a band gap at DP. This behavior can be understood in terms of Bragg scattering of electron waves on the disorder potential due to the adatoms. In the ordered state,  some of the harmonics of the disorder potential scatter electrons coherently, providing a gap-opening perturbation. Such harmonics cease to contribute to the incoherent scattering responsible for momentum relaxation. This leads to an increase in the mean free path, and in conductivity.

\section{T-matrix of a single scatterer}

When the adatoms and groups such as H, F and OH bind to a carbon atom in graphene lattice, one of the $\pi$ electrons turns into a $\sigma$ electron, thereby being removed from the conduction band. As a result, the site of the functionalized carbon atom 
is seen by other $\pi$ electrons as a vacancy.
Simultaneously, a quasibound electron state apears, giving rise to a low-energy resonance in the local density of states near the adatom. {\it Ab initio} study \cite{Wehling08} predicts the resonance energy values which span a wide range:
$\epsilon_*=-0.03,\,-0.11,\,-0.70,\,-0.67\,{\rm eV}$ for H, ${\rm CH_3}$, OH and F, respectively. The resonance at $\epsilon=\epsilon_*$ manifests itself as a resonance peak in the electron scattering amplitude.

A simplest model for electron scattering on point-like defects is provided by a delta-function potential, described by the Hamiltonian
\be\label{eq:Hamiltonian}
H=\sum_{|\vec x-\vec x'|=1} t_0 (\psi^{\dagger}_{\bf x}\psi_{\bf x'}+{\rm h.c.})+\sum_{\bf x} u({\bf x})\psi^{\dagger}_{\bf x} \psi_{\bf x}
,
\ee
with the hopping amplitude $t_0\approx 3.1\, {\rm eV}$, and potential $u(\vec x)$ equals $U$ on the site occupied by an adatom and zero elsewhere.

The regime of weak scattering, described by small $U\ll t_0$, can be easily treated in the Born approximation (see e.g. Refs.[\onlinecite{Ostrovsky06,Nomura06}]). Here we are interested in the regime of strong scattering, $U\gg t_0$.
Physically, such strong scattering potential corresponds to effectively removing an electronic level from the carbon site occupied by an adatom, by prohibiting electrons to visit it. Scattering on a strong, point-like defect can be conveniently described by a T-matrix, defined via the Greens function for multiple scattering. In the operator form, $\hat T=\hat U+\hat T G\hat U$, where $\hat U$ is the defect potential and $G$ is the free particle Greens function.

Direct calculation shows that the behavior of the T-matrix for the tight-binding model (\ref{eq:Hamiltonian}) is the same as that found in the continual  massless Dirac particle model with two degenerate valleys \cite{Pepin97}. Treating the defect as a delta-function potential with equal inter-valley and intra-valley scattering amplitudes, we find
\be\label{eq:T}
T(\epsilon)=\frac{\pi v_0^2}{\delta+\frac{\epsilon}{2}\ln (-W^2/\epsilon^2)}
,\quad
\delta=\frac{\pi v_0^2}{U}n_0
\ee
where $W\approx 3t_0$, $n_0=\frac{2}{3\sqrt{3} a^2}$ is the density of carbon atoms in one sublattice (thus $2n_0$ is the net density of carbon atoms in graphene lattice). 

The $T$-matrix (\ref{eq:T}) has a resonant form, with a complex pole at an energy $\epsilon_*$ determined from the equation 
$
\delta+\frac{1}{2}\epsilon_* \ln(- W^2/\epsilon_*^2)=0$. 
Since $\delta$ is small for a strong scatterer, $U\gg t_0$, the energy $\epsilon_*$ is approximately given by
\be\label{eq:epsilon_*}
\epsilon_*\approx -\frac{\delta}{\ln (W/|\delta|)- i\frac{\pi}2 {\rm sgn \delta}}
. 
\ee
The resonance (\ref{eq:epsilon_*}) is positioned at a low energy, as expected from the equivalence between the tight-binding and continual models. 
The energy $\epsilon_*$ vanishes, when the adatom potential is infinitely strong, $U\to \infty$, turning into a zero mode associated with a vacancy \cite{Pereira06,Pereira08}. 

In our resonant scattering model, the parameters $U$ and $\delta$ appear as primary quantities which are used to derive the resonance energy $\epsilon_*$. However, since the energy value $\epsilon_*$ is provided by {ab initio} calculations, we use its relation with the parameters $\delta$ and $U$ to fix their values in order to generate a correct value of $\epsilon_*$. 

While these results seem reasonable, one may recognize that potential strength exceeding the bandwidth parameter $t_0$ is unphysical, and seek
an alternative description. In Refs.[\onlinecite{Wehling08,Robinson08}] the defects arising as a result of $sp^2\to sp^3$ hybridization were modelled using a localized level hybridized with the $\pi$-band of graphene. One can see by a direct calculation that the T-matrix in this case has the same behavior at low energies as that found for a point-like scatterer, Eq.(\ref{eq:T}). 

The agreement between the two models is by no means a coincidence. On very general grounds, we can start from the unitarity theorem for the T-matrix of a point-like scatterer, 
\be\label{eq:unitarity}
T_0^*(\epsilon)-T_0(\epsilon)=i\pi \frac{|\epsilon|}{\pi v_0^2} |T_0(\epsilon)|^2.
\ee
Rewriting this relation as $T_0^{-1}(\epsilon)-{T_0^{-1}}^*(\epsilon)=i\pi \frac{|\epsilon|}{\pi v_0^2}$, and taking into account that $T_0(\epsilon)$ is analytic in the upper halfplane $\Im\epsilon>0$, we find
\be\label{eq:T_0}
T_0(\epsilon)=\frac{2\pi v_0^2}{f(\epsilon)+{\epsilon} \ln\left(-\frac{W^2}{\epsilon^2} \right)},
\ee
where $W$ is the band width, and the function $f$ must be purely real for the T-matrix to satisfy the unitarity relation. 

In our calculation, $T_0(\epsilon)$ will enter the retarded Greens function. Accordingly, in the above equation we choose the branch of the logarithm that is analytic in the upper half-plane, and is real on the imaginary axis. The $T$-matrix which is analytic in the lower half-plane, obtained by complex conjugation, is given by $T_0^*(\epsilon)$. This quantity will enter the advanced Greens function.

The above form of the $T$-matrix, Eq.(\ref{eq:T_0}), is quite general. It includes the $T$-matrices obtained in the two models proposed for description of adatoms -- the local potential model, Eq.(\ref{eq:Hamiltonian}), and the model of a localized level hybridized with the $\pi$-band of graphene \cite{Wehling08,Robinson08}. The local potential model, described above,  corresponds to $f(\epsilon)=2\delta$, where $\delta$ is defined in Eq.(\ref{eq:T}).

\section{The self-consistent T-matrix approximation}

The T-matrix approach can be used to describe electron scattering in the presence of an ensemble of adatoms. Generalizing our discussion of a single scatterer, we model the system by the Hamiltonian (\ref{eq:Hamiltonian}) with the potential 
\be
u(\vec x)=\sum_i U\delta(\vec x-\vec x_i)
\ee
which equals $U$ on the sites occupied by adatoms, and zero elsewhere. We shall assume that the adatoms populate the sublattices $A$ and $B$ unequally, with concentrations  $n_A$ and $n_B$, while the distribution within each sublattice is completely random (see Ref.[\onlinecite{Abanin10}]). The concentrations $n_A$ and $n_B$ are taken to be equal in the disordered state and are unequal in the ordered state.

To make connection with previous work, we note that the dimensionless occupancies $0\le n_A\le 1$ and $0\le n_B\le 1$ used in Ref.[\onlinecite{Abanin10}] correspond to $n_A/n_0$ and $n_B/n_0$ in the present notation. The dimensionfull concentrations are denoted $\tilde{n}_{A}$ and $\tilde{n}_{B}$ in Ref.[\onlinecite{Abanin10}].

Within the SCTA approach, the collective effect of disorder on electron scattering can be described by the disorder-averaged Greens functions and self-energies which satsify a selfconsistency relation \cite{Ostrovsky06}. In our case, the disorder-averaged Greens function for each valley takes the following form
\be\label{eq:G}
G(\epsilon,{\bf k})=\frac1{\epsilon-\Sigma-v_0{\bf \sigma k}}
,\quad 
\Sigma= \left[\begin{array}{cc}
         \Sigma_A & 0\\
         0&  \Sigma_B
      \end{array}
 \right].
\ee
where $\sigma=(\sigma_x,\sigma_y)$ are the Pauli matrices.

The self-energies are obtained by averaging the T-matrices over the position of impurities, separately for the $A$ and $B$ sublattices,
\be\label{eq:Sigma1}
\Sigma_{A}=n_{A} \la T_{A} \ra
,\quad
\Sigma_{B}=n_{B} \la T_{B} \ra
.
\ee
By a direct calculation we express 
the disorder-averaged $T$-matrices in Eq.(\ref{eq:Sigma1}) through the self-energies:
\bea\label{eq:T_averaged_A}
&& \la T_{A} \ra=\frac{\pi v_0^2}{f(\epsilon)+\frac{\epsilon_{B}}{2} \ln\left(-\frac{W^2}{\epsilon_A \epsilon_B} \right)},
\\\label{eq:T_averaged_B}
&& \la T_{B} \ra=\frac{\pi v_0^2}{f(\epsilon)+\frac{\epsilon_{A}}{2} \ln\left(-\frac{W^2}{\epsilon_A \epsilon_B} \right)},
\eea
where the quantities $\epsilon_A$, $\epsilon_B$ are defined as
\be\label{eq:epsilon_AB}
\epsilon_A=\epsilon-\Sigma_A
,\quad
\epsilon_B=\epsilon-\Sigma_B
.
\ee
The branch of the logarithm in Eqs.(\ref{eq:T_averaged_A}),(\ref{eq:T_averaged_B}) is defined in the same way as in the definiton of the $T$-matrix for a single scatterer, Eq.(\ref{eq:T_0}). 

The analogues of Eqs.(\ref{eq:Sigma1}),(\ref{eq:T_averaged_A}),(\ref{eq:T_averaged_B}) for the advanced Greens function can be obtained by replacing the self-energies $\Sigma_{A(B)}$ and the quantities $\epsilon_{A(B)}$ by their complex conjugates, $\Sigma^*_{A(B)}$ and $\epsilon^*_{A(B)}=\epsilon-\Sigma^*_{A(B)}$. 

The equations for self-energies and can be solved numerically, and, in the limit of high density, analytically (see next section). 
The disorder-averaged electronic properties of the system, such as density of states and conductivity, can be determined from quantities $\epsilon_{A},\epsilon_{B}$. 

\section{Calculation of conductivity}

We seak to evaluate the conductivity of graphene as a function of the adatom concentrations on the two sublattices, $n_A$ and $n_B$. We will use a general expression for conductivity, obtained from SCTA, in which we perform virial expansion in powers of $n_A$ and $n_B$. The leading (Drude) contribution is inversely proportional to the total concentration $n_A+n_B$, and is thus insensitive to the imbalance of $n_A$ and $n_B$. In subleading order, we find a contribution to conductivity which depends on the sublattice imbalance as $\Delta\sigma\propto n_An_B/(n_A+n_B)^2$. This contribution is negative, which makes the conductivity of the ordered state, $n_A\ne n_B$, greater than that of the disordered state, $n_A=n_B$.

We start with a general expression for conductivity \cite{Ostrovsky06},
\be\label{eq:conductivity}
\sigma(\epsilon)=\frac{1}\pi e^2 v_0^2 \Pi _{RA} (\epsilon),
\ee
where $\epsilon$ is the Fermi energy, and
\be\label{eq:piRA}
\Pi_{RA} (\epsilon)={\rm Tr}\int \frac{d^2k}{(2\pi)^2} G^{R} (\epsilon,{\bf k})G^{A}(\epsilon,{\bf k}).
\ee
Here the integration is performed over the whole Brillouin zone, and trace is taken over sublattice indices. 

Using the disorder-averaged Greens function (\ref{eq:G}), and evaluating the integral over $\vec k$, we rewrite 
\be\label{eq:PiRA1}
\Pi^{RA}(\epsilon)=\frac{1}{2\pi v_0^2} \frac{\epsilon_A \epsilon^*_A+ \epsilon_B \epsilon^* _B}{\epsilon_A\epsilon_B-\epsilon^*_A \epsilon^*_B}(F(\epsilon_A\epsilon_B)-{F^*}(\epsilon_A\epsilon_B)),
\ee
where $F=\ln\left( -\frac{W^2}{\epsilon_A\epsilon_B}\right)$, 
$F^*=\ln\left( -\frac{W^2}{\epsilon_A^*\epsilon_B^*}\right)$,
with the logarithm branch chosen to be analytic in the upper (lower) half-plane.
In particular, for real $\epsilon_A=\epsilon_B=\epsilon$, we have $F=\ln\left|\frac{W^2}{\epsilon^2} \right | + i \pi {\rm sgn}(\epsilon)$, $F^*=\ln\left|\frac{W^2}{\epsilon^2} \right | - i \pi {\rm sgn}(\epsilon)$; we will use this relation below.

Eqs.(\ref{eq:conductivity}),(\ref{eq:PiRA1}) give a general answer for conductivity in terms of self-energies. Performing virial expansion of $\Sigma_A$ and $\Sigma_B$ in $n_A$ and $n_B$, we write
\bea\nonumber
\Sigma_A &=&
n_A \la T_A\ra -n_A^2\la T_A\ra\frac{\p \la T_A\ra}{\p\epsilon_A}
\\\label{eq:Taylor1}
&& -n_An_B\la T_B\ra\frac{\p \la T_A\ra}{\p\epsilon_B}+...
\\ \nonumber
\Sigma_B &=& n_B \la T_B\ra -n_B^2\la T_B\ra\frac{\p \la T_B\ra}{\p\epsilon_B}
\\\label{eq:Taylor2}
&& -n_An_B\la T_A\ra\frac{\p \la T_B\ra}{\p\epsilon_A}+...
\eea
where the T-matrices $\la T_A\ra$ and $\la T_B\ra$ and their derivatives on the right hand side are evaluated at $n_A=n_B=0$, as appropriate for Taylor expansion of Eqs.(\ref{eq:T_averaged_A}),(\ref{eq:T_averaged_B}).

At leading order, keeping only the terms of first order in $n_A$ and $n_B$, we find 
\be\label{eq:PiRA0}
\Pi^{RA}_0(\epsilon)=\frac{2}{v_0^2}\frac{1}{n_A+n_B}\frac{i|\epsilon|}{T_0^*(\epsilon)-T_0(\epsilon)}.
\ee
This expression can be simplified using the unitarity relation, Eq.(\ref{eq:unitarity}), to obtain
\be\label{eq:sigma_0}
\sigma_0(\epsilon)=\frac{2}{\pi} \frac{e^2 v_0^2}{|T_0(\epsilon)|^2(n_A+n_B)}.
\ee
This expression, which is inversely proportional to the total adatom concentration $n_A+n_B$, is nothing but the Drude conductivity. Being a function of the total concentration, this contribution is not sensitive to sublattice imbalance.

For energies $\epsilon$ sufficiently far from DP, $|\epsilon|\gg |\epsilon_*|$, we can approximate $T(\epsilon)\approx \pi v_0^2/(\epsilon\ln(iW/\epsilon))$.
In this limit, Eq.(\ref{eq:sigma_0}) becomes
\be\label{eq:sigma_simple}
\sigma_0(\epsilon)=\frac{e^2}{2\pi^3  v_0^2}\frac{\epsilon^2 \ln ^2(W^2/\epsilon^2)}{n_A+n_B}
,
\ee
which is the result found in Ref.[\onlinecite{Ostrovsky06}].

The expansion in powers of $n_A$ and $n_B$, employed in this and next sections, is controlled by the parameter 
\be
n_{A(B)}|\p T/\p\epsilon|\ll 1.
\ee 
Evaluating the derivative, $\p T/\p\epsilon=\pi v_0^2\ln(iW/\epsilon)/(\delta+\epsilon\ln(iW/\epsilon))^2$, we can write this condition as
\be
\frac{n_{A(B)}}{\pi v_0^2}|T_0(\epsilon)|^2\ln\frac{W}{|\epsilon|}\ll 1
.
\ee
This condition can also be written in terms of the mean free path, see Eq.(\ref{eq:k_Fl_log}).
Recalling that Drude conductivity in two dimensions is of order $k_F\ell\, e^2/h$, and comparing to the result (\ref{eq:sigma_0}), we obtain Eq.(\ref{eq:k_Fl_log}). Thus, notably, the parameter controlling our virial expansion is $k_F\ell/\ln(W/|\epsilon|)\gg 1$
rather than $k_F\ell\gg 1$. 

\section{The contribution $\Delta\sigma$}

The subleading terms in the expansion of conductivity in powers of adatom concentration are zero-order in $n_A$ and $n_B$. According to Eq.(\ref{eq:k_Fl_log}), such terms are also zero-order in $k_F\ell$ and are of order $e^2/h$. As we shall see, there are two distinct contributions to conductivitiy arising at this order. One is a contribution proportional to $n_An_B/(n_A+n_B)^2$, which we shall denote as $\Delta \sigma$. This contribution depends on the imbalance $n_A/n_B$ and thus describes the effect of sublattice ordering on conductivity. In addition, there is a weak localization contribution $\delta\sigma_{WL}$ which cannot be obtained from SCTA. However, as we argue below, to leading order this contribution is independent of $n_A$ and $n_B$ and thus is inessential.

We proceed with Taylor expanding Eqs.(\ref{eq:Sigma1}),(\ref{eq:T_averaged_A}),(\ref{eq:T_averaged_B}) to second order in $n_A$ and $n_B$ as outlined above. In order to extract the contribution that depends on the sublattice imbalance $n_A/n_B$, we rearrange the quadratic terms in Eq.(\ref{eq:Taylor1}) as follows:
\bea\nonumber
&& n_A^2\la T_A\ra\frac{\p \la T_A\ra}{\p\epsilon_A}+n_An_B\la T_B\ra\frac{\p \la T_A\ra}{\p\epsilon_B}=n_A(n_A+n_B)
\\\nonumber
&& \times \la T_A\ra\frac{\p \la T_A\ra}{\p\epsilon_A}
+n_An_B\lp \la T_B\ra\frac{\p \la T_A\ra}{\p\epsilon_B}
-\la T_A\ra\frac{\p \la T_A\ra}{\p\epsilon_A}\rp
,
\eea
and similarly in Eq.(\ref{eq:Taylor2}).
The terms proportional to $n_A+n_B$ can be dropped since they cannot produce a contribution that depends on the imbalance. Evaluating the derivatives in $\epsilon_A$ and $\epsilon_B$, we note that the nonzero answer arises solely due to the prefactors $\epsilon_A$ and $\epsilon_B$ of the log terms in Eqs.(\ref{eq:T_averaged_A}),(\ref{eq:T_averaged_B}), since the contributions arising from differentiating $\ln(-W^2/\epsilon_A\epsilon_B)$ cancel each other. 
Keeping only the terms proportional to $n_An_B$, we write
\bea\label{eq:Sigma_expand}
&&\Sigma_{A(B)}=n_{A(B)} T_0(\epsilon)+\frac{T_0^3(\epsilon)}{2\pi v_0^2}n_A n_B  F(\epsilon),
\\\label{eq:Sigma_expand2}
&&\Sigma^*_{A(B)}=n_{A(B)} T_0^*(\epsilon)+\frac{{T_0^*}^3(\epsilon)}{2\pi v_0^2}n_A n_B {F^*}(\epsilon)
,
\eea
These expressions are then plugged into Eq.(\ref{eq:PiRA1}), and expanded to keep the two lowest orders in $n_A$, $n_B$. This gives
\be\label{eq:PiRA2}
\Pi^{RA}(\epsilon)=\Pi^{RA}_0(\epsilon)+\Pi^{RA}_1(\epsilon).
\ee
where $\Pi^{RA}_0(\epsilon)$ is given by Eq.(\ref{eq:PiRA0}). The contribution $\Pi^{RA}_1$ can be represented as a sum of two terms,
\be\label{eq:PiRA_order1}
\Pi^{RA}_1(\epsilon)=\tilde\Pi^{RA}_1(\epsilon)+\delta\Pi^{RA}_1 (\epsilon),
\ee
where $\tilde\Pi^{RA}_1$ depends on $n_A$ and $n_B$, while $\delta\Pi^{RA}_1$ does not. The latter term, which is given by
\be\label{eq:PiRA_1_indep}
\delta\Pi^{RA}_1=
-\frac{1}{\pi v_0^2}-\frac{i{\rm sgn}\epsilon}{v_0^2}\frac{T_0(\epsilon)+T_0^*(\epsilon)}{T_0^*(\epsilon)-T_0(\epsilon)},
\ee
will be ignored in what follows.

The contribution $\tilde\Pi^{RA}_1$, after some algebra, can be written as
\begin{widetext}
\be\label{eq:tildePiRA}
\tilde\Pi^{RA}_1=-\frac{2i|\epsilon|}{v_0^2(T_0^*(\epsilon)-T_0(\epsilon))^2}\frac{{n}_A {n}_B}{({n}_A+{n}_B)^2}
\left[\frac{T_0^2(\epsilon)-{T_0^*}^2(\epsilon)}{\epsilon} +\frac{\ln (W^2/\epsilon^2)}{\pi v_0^2} ({T_0^*}^3(\epsilon)-T_0^3(\epsilon))-\frac{i{\rm sgn}\epsilon}{v_0^2}({T_0^*}^3(\epsilon)+T_0^3(\epsilon)) \right].
\ee
\end{widetext}
The three terms in this expression originate from an expansion of the denominator in Eq.(\ref{eq:PiRA1}) in  $n_A$, $n_B$ as follows: the first term comes from the products in $\Sigma_A\Sigma_B-\Sigma^*_A\Sigma^*_B$, where self-energies are taken in the lowest order in $n_A$, $n_B$; the second and third terms come from the term $-\epsilon (\Sigma_A+\Sigma_B-\Sigma^*_A-\Sigma^*_B)$, where the self-energies are taken in the second order in $n_A$, $n_B$.

\begin{figure}
\includegraphics[width=3.4in]{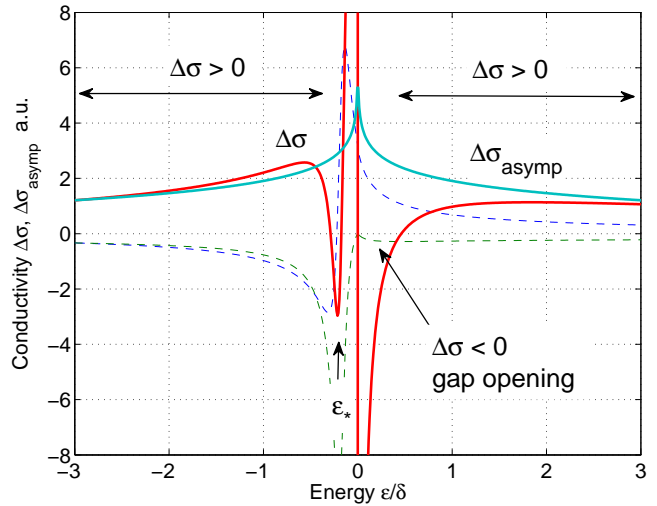}
\vspace{-4mm}
\caption[]{Energy dependence of the contribution to conductivity which is sensitive to sublattice ordering. Shown is the contribution obtained by virial expansion in $n_A$, $n_B$, Eq.(\ref{eq:tildePiRA2}), and its high-energy asymptotic form, Eq.(\ref{eq:delta_sigma}), for $W/\delta=20$. Real and imaginary parts of a single-scatterer T-matrix, Eq.(\ref{eq:T}), are shown by dashed lines. The virial expansion is valid at densities detuned far enough from DP, Eq.(\ref{eq:k_Fl_log}). Conductivity increases due to sublattice ordering at energies away from DP and is reduced at energies $\epsilon>0$ near DP, in the region where an energy gap opens up in the ordered state.
}
\vspace{-4mm}
\label{fig1}
\end{figure}

The result (\ref{eq:tildePiRA}) can be simplified using the unitarity relation for the $T$-matrix, Eq.(\ref{eq:unitarity}). Subtracting the value at $n_A=n_B$, and using Eq.(\ref{eq:conductivity}), gives a contribution describing the change in conductivity due to ordering,
\begin{widetext}
\be\label{eq:tildePiRA2}
\Delta\sigma=\frac{e^2v_0^2}{2\pi|T_0(\epsilon)|^2}\frac{({n}_A -{n}_B)^2}{({n}_A+{n}_B)^2}
\left[-\frac{T_0(\epsilon)+{T_0^*}(\epsilon)}{\epsilon} +\frac{\ln (W^2/\epsilon^2)}{\pi v_0^2} ({T_0^*}^2(\epsilon)+T_0^*(\epsilon)T_0(\epsilon)+T_0^2(\epsilon))-\frac{{T_0^*}^3(\epsilon)+T_0^3(\epsilon)}{\epsilon |T_0(\epsilon)|^2} \right].
\ee
\end{widetext}
Energy dependence of $\Delta\sigma$ is shown in Fig.\ref{fig1}. 
The contribution (\ref{eq:tildePiRA2}) is positive sufficiently far from DP, $|\epsilon|\gg |\epsilon_*|$, and negative near DP at $\epsilon>0$.

The expression (\ref{eq:tildePiRA2}) simplifies at large detuning from DP, since 
at large $\epsilon$ the imaginary part of $T_0(\epsilon)$ is small compared to its real part. Thus, in Eq.(\ref{eq:tildePiRA2}) we can approximate $T_0(\epsilon)\approx T_0^*(\epsilon)\approx \Re  T_0(\epsilon)$. This gives 
\bea\label{eq:tildePiRA3}
&& 
\Delta\sigma\approx \frac{e^2v_0^2}{2\pi (\Re T_0(\epsilon))^2}\frac{(n_A -n_B)^2}{(n_A+n_B)^2}
\\ \nonumber
&&\times \left[  -4\frac{\Re  T_0(\epsilon)}{\epsilon} +\frac{3}{\pi v_0^2} (\Re  T_0(\epsilon))^2 \ln (W^2/\epsilon^2) \right].
\eea
Replacing $\Re T_0(\epsilon)$ by $\pi v_0^2/(\epsilon\ln(W/|\epsilon|))$, we obtain the desired contribution to the conductivity, Eq.(\ref{eq:delta_sigma}).

Taking the ratio of $\Delta\sigma$ and the Drude conductivity, Eqs.(\ref{eq:delta_sigma}),(\ref{eq:sigma_simple}), we have
\be
\frac{\Delta\sigma}{\sigma_0}\sim\frac{\ln(W/|\epsilon|)}{k_F\ell}
,
\ee
where we used the relation $\sigma_0\sim k_F\ell\, e^2/h$. Comparing to Eq.(\ref{eq:k_Fl_log}), we see that $\Delta\sigma/\sigma_0\ll1$ in the regime of validity of our virial expansion.

It is interesting to compare the contribution $\Delta\sigma$, Eq.(\ref{eq:delta_sigma}), to the weak localization (WL) contribution, which arises in the same order of an expansion in $1/(k_F\ell)$.
The WL contribution is given by \cite{Ostrovsky06}
\be\label{eq:wl}
\delta\sigma_{WL}=-\frac{e^2}{\pi^2}\ln  \frac{L}{\ell},
\ee
where $L$ is the sample size. We note that at leading order in $1/(k_F\ell)$ the mean free path $\ell$ depends only on the total adatom concentration $n_A+n_B$, but not on the sublattice imbalance. This can be understood by combining the result for Drude conductivity, Eq.(\ref{eq:sigma_simple}), with the relation $k_F\ell=\sigma h/e^2$. We obtain, to leading order in $k_F\ell$,  
\be
\ell_0=\frac{|\epsilon|}{\pi^2 v_0 (n_A+n_B)}\ln^2 (W^2/\epsilon^2)
,
\ee 
a result which is independent of the sublattice imbalance. 

The subleading term in the expansion in powers of $k_F\ell$, which is of order $\delta\ell\sim \ell/(k_F\ell)=\frac{1}{k_F }$, may depend on the imbalance $n_A/n_B$. However, the corresponding contribution to the $\delta\sigma_{WL}$, Eq.(\ref{eq:wl}), arises only from $\ell$ under the log, and thus is of order $1/(k_F\ell)$ in units $e^2/h$. This is parametrically smaller than the contribution $\Delta\sigma$, Eq.(\ref{eq:delta_sigma}). From this we conclude that the effect of sublattice imbalance arising from the WL contribution is inessential compared to the $\Delta\sigma$ contribution.

\section{Summary and discussion}

In summary, sublattice ordering of defects directly manifests itself in transport properties. Conductivity increases in the metallic regime far from DP, because in the ordered state part of the scattering ceases to contribute to momentum relaxation, instead acquiring a coherent, Bragg-like character. Simultaneously, in the ordered state, conductivity is reduced at energies near DP, where the band gap opens up\cite{Abanin10}.

These effects were investigated using a virial expansion of conductivity in the adatom concentration, a method that is valid in the metallic state at sufficiently large mean free path, Eq.(\ref{eq:k_Fl_log}). One can expect transport in the state with a well developed gap to be dominated by the low-energy modes propagating along domain boundaries, separating regions with opposite signs of the order parameter (sublattice imbalance), as well as by thermal activation across the gap. Theoretical description of this regime lies outside the scope of this paper.

Finally, we note that the results of this work apply to other scenarios of partial ordering of defects, in which populations on several sublattices become spontaneously imbalanced, the distribution on each sublattice remaining random. One example is the recently discussed Kekule-type ordering\cite{Cheianov09} in which three equivalent sublattices become unequally populated. We expect that the picture of Bragg scattering on the density harmonics arising due to ordering, accompanied by a reduction in disorder scattering and an increase in conductivity, applies in this case. 
An increase in conductivity can serve as an experimental signature of a transition to an ordered state.

\section{Acknowledgements}

We thank A.V. Shytov for useful discussions. DA thanks Aspen Center for Physics where part of this work was completed. This work was supported by Office of
Naval Research Grant No. N00014-09-1-0724 (LL).

\end{document}